\documentclass[aps,a4paper,floatfix, nofootinbib]{revtex4}

\usepackage[utf8]{inputenc}

\usepackage{graphicx,color}
\usepackage{amsmath,amssymb,amsfonts,amsthm,amscd,bm}
\usepackage{booktabs}
\usepackage{cancel}

\def\*{\star}
\def\[{\left[}
\def\]{\right]}
\def\({\left(}      

\def\){\right)}

\def\frac#1#2{\dfrac{#1}{#2}}
\def\inv#1{\dfrac{1}{#1}}

\def\d{\partial}

\def\2pi{\hbox{$2\pi i$}}

\def\dsl{\raise.15ex\hbox{/}\kern-.57em\partial}
\def\Dsl{\,\raise.15ex\hbox{/}\mkern-.13.5mu D}
   \def\CE{{\cal E}}   \def\CF{{\cal F}}
      
      \def\CL{{\cal L}}
      \def\CO{{\cal O}}
      
\def\CS{{\cal S}}      
\def\CV{{\cal V}}      
   \def\CZ{{\cal Z}}

\def\2pi{\hbox{$2\pi i$}}

\def\dsl{\raise.15ex\hbox{/}\kern-.57em\partial}
\def\Dsl{\,\raise.15ex\hbox{/}\mkern-.13.5mu D}
\font\numbers=cmss12
\font\upright=cmu10 scaled\magstep1
\def\stroke{\vrule height8pt width0.4pt depth-0.1pt}
\def\topfleck{\vrule height8pt width0.5pt depth-5.9pt}
\def\botfleck{\vrule height2pt width0.5pt depth0.1pt}
\def\Zmath{\vcenter{\hbox{\numbers\rlap{\rlap{Z}\kern
    0.8pt\topfleck}\kern 2.2pt
    \rlap Z\kern 6pt\botfleck\kern 1pt}}}
\def\Qmath{
    \vcenter{\hbox{\upright\rlap{\rlap{Q}\kern3.8pt\stroke}\phantom{Q}}}}
\def\Nmath{\vcenter{\hbox{\upright\rlap{I}\kern 1.7pt N}}}
\def\Cmath{\vcenter{\hbox{\upright\rlap{\rlap{C}\kern
                   3.8pt\stroke}\phantom{C}}}}
\def\Rmath{\vcenter{\hbox{\upright\rlap{I}\kern 1.7pt R}}}
\def\Z{\ifmmode\Zmath\else$\Zmath$\fi}
\def\Q{\ifmmode\Qmath\else$\Qmath$\fi}
\def\N{\ifmmode\Nmath\else$\Nmath$\fi}
\def\C{\ifmmode\Cmath\else$\Cmath$\fi}
\def\R{\ifmmode\Rmath\else$\Rmath$\fi}
\def\barray{\begin{eqnarray}}
\def\earray{\end{eqnarray}}
\def\beq{\begin{equation}}
\def\eeq{\end{equation}}

\def\n{\noindent}
\def\kvec{{\bf{k}}}

\def\smallhalf{{\scriptstyle \inv{2}}}

\def\smallhalf{{\textstyle \inv{2}}}

\def\AA{\leavevmode\setbox0=\hbox{h}
\dimen0=\ht0 \advance\dimen0 by-1ex\rlap{\raise.67\dimen0\hbox{\char'27}}A}

\def\blue#1{{\color{blue}{#1}}}

\makeatletter
\def\iddots{\mathinner{\mkern1mu\raise\p@
\vbox{\kern7\p@\hbox{.}}\mkern2mu
\raise4\p@\hbox{.}\mkern2mu\raise7\p@\hbox{.}\mkern1mu}}
\makeatother

\def\dim#1{ {\bf [} #1 {\bf ]}  }

\theoremstyle{plain}

\theoremstyle{remark}

\begin{document}

\def\cstarUV{c^*_{UV}}

\def\kvec{{\bf k}}

\def\rhovac{\rho_{\rm vac}}
\def\smallhalf{\tfrac{1}{2}}

\def\gcal{\mathfrak{g} }

\def\cIR{c_{\rm IR}}
\def\cUV{c_{\rm UV}}

\def\cftIR{{\rm cft}_{\rm IR}}
\def\cftUV{ {\rm cft}_{\rm UV}}

\def\cbulk{c_{\rm bulk}}

\def\ceff{c_{\rm eff}}

\def\cpert{c_{\rm pert}}

 \def\GammaUV{\Gamma_{\rm UV}}

\def\vol{\CV} 

\def\phinot{\phi_0}

\def\lambdahat{\mu}

 \def\GammaUV{\Gamma_{\rm UV}}

 \def\Gammarho{\Gamma_\rho}

\title{
Comment  on the vacuum energy density  for $\lambda \phi^4$ theory in $d$ spacetime dimensions
}
\author{
 Andr\'e  LeClair\footnote{andre.leclair@gmail.com}
}
\affiliation{Cornell University, Physics Department, Ithaca, NY 14853, United States of America} 

\begin{abstract}

In a recent article we showed that the vacuum energy density  in two spacetime dimensions for a wide variety of  integrable quantum field theories 
has the form $\rhovac = - m^2 /2 \gcal$ where $m$ is a physical mass and $\gcal$ is a generalized coupling,   where in the free field limit $\gcal \to 0$,  $\rhovac$ diverges.  
This vacuum energy density has the form $\langle T_{\mu\nu}  \rangle = - \rhovac g_{\mu\nu}$,   and has previously been considered as a contribution to the stress energy tensor in Einstein's gravity as a 
``cosmological constant".    
We speculated that in four spacetime dimensions $\rhovac$ takes a similar form $\rhovac = - m^4/2 \gcal$,   but did not support this idea in any specific model.  
In this article we study this problem for $\lambda \phi^4$ theory in $d$ spacetime dimensions.     We show how to obtain the {\it exact } $\rhovac$ for the sinh-Gordon theory  in the weak coupling limit by using a saddle point approximation.    This calculation indicates that the vacuum energy can be well-defined,   positive or negative,  without 
spontaneous symmetry breaking.      We also show that $\rhovac$ satisfies a Callan-Symanzik type of renormalization group equation.    
For the most interesting case physically,  $\rhovac$ is positive and can arise from a marginally relevant negative coupling $\gcal$ and the vacuum energy  flows to zero at low energies.   

\end{abstract}

\maketitle
\tableofcontents

\section{Introduction}

The so-called "cosmological constant problem"   continues to provide serious challenges to our understanding of fundamental physics.    
    Einstein's equations of general relativity involve the classical stress-energy tensor as a source of gravitation  and should include all possible sources of stress energy. 
Experimental cosmology provides evidence for a very small positive cosmological constant,  and the origins of it remain unknown.   There are many possibilities that remain to be explored,  everything from modified classical General Relativity,   quantum fluctuations of the vacuum,  to quantum gravity effects at the Planck scale.     
In this article we study this problem from the point of view  in which it was first stated,  namely as originating in  quantum vacuum fluctuations,   which is where the often 
quoted discrepancy by 120 orders of magnitude originated.  However we do not claim any kind of resolution of the problem,   since the non-zero cosmological constant may have completely different origins.    Nevertheless,   it is worthwhile to fully explore this option.   
We need to say however that some researchers believe that this problem cannot be resolved without considering quantum fields in curved spacetime,  which is far beyond the scope of this article.   See for instance the recent articles \cite{Peracaula,Peracaula2} and references therein.

 In a semi-classical quantum theory  it is reasonable to suppose that the classical $T_{\mu\nu}$  is replaced by its quantum vacuum expectation value $ \langle 0 | T_{\mu\nu}  | 0 \rangle$, where $|0\rangle$ is the vacuum state.
Based on general coordinate invariance one expects
\beq
\label{CC1}
\langle 0 | T_{\mu\nu}  | 0 \rangle = - \rhovac \, g_{\mu\nu}
\eeq
where $g_{\mu\nu}$ is the spacetime metric.   In the above equation the convention for the metric is  the signature  
$g_{\mu \nu} = {\rm diag} ( -1, 1,1,1)$,  i.e. $g_{00} =  -g_{ii} = -1$ in Minkowski space.     
Perhaps the first version of the cosmological constant problem was based on viewing 
a {\it free}  quantum field 
 as  a collection of harmonic oscillators of frequency $\omega_k = \sqrt{\kvec^2 + m^2}$,  and the vacuum energy is naively the sum of the zero point energies \cite{Weinberg,Martin}:   
\beq
\label{CC2}
\rhovac =  \int_0^\Lambda \frac{dk}{(2 \pi )^3 } \,  4 \pi k^2   ~ \smallhalf  \sqrt{k^2  + m^2}  \approx   \frac{\Lambda^4}{16 \pi^2}  
\eeq
where $\Lambda$ is an ultraviolet cutoff and we have assumed $\Lambda\gg m$. 
The problem  is that for reasonable values of the cut-off $\Lambda$,   such as the Planck scale,   the above $\rhovac$ is off  by roughly 120  orders of magnitude compared to 
astrophysical measurements.     The original problem has evolved to consider a series of phase transitions in the thermal development of the dynamical evolution of the Universe 
where $\Lambda$ is a scale of spontaneous symmetry breaking (SSB),   such as the electro-weak scale,   a supersymmetry breaking scale,   or even the  QCD scale  (see for example the review \cite{Carroll} and references therein.)   
In any case,  the corresponding $\Lambda$ leads to much too high a scale to explain the observed astrophysical value of $\rhovac$.    

We emphasize that we will not consider  here quantum fields in curved spacetime,  nor quantum gravity,   in fact we neglect gravity entirely.  This is the main shortcoming of this article,  since as previously mentioned,   some physicists feel that one cannot avoid dealing with gravity in order to even attempt to solve the cosmological constant problem.    Although ignoring gravity  may turn out to be an oversimplification,   let us mention that one does not need to understand quantum electrodynamics in curved spacetime in order to understand the cosmic microwave background,
so as a first step we can suppose this is true for the cosmological constant itself.   
We feel this is justified  since the original version of the cosmological constant problem was based on divergences in the vacuum energy in pure quantum field theory without gravity,  
and it is worthwhile to make sense of this  since it is not well understood;   it is worth exploring whether it can be ruled out or not,   and we will argue that it is not.
Although perhaps an abuse of terminology,   henceforth  we refer to   ``$\rhovac$"   and  ``cosmological constant" interchangeably,  although clearly we are not studying cosmology per se but rather the vacuum energy density of an interacting QFT in flat space.

One should strongly question the above naive computation in \eqref{CC2},    since we are accustomed to dealing with divergences in quantum field theory  (QFT) in a way that leads to finite
physical predictions.    Also,   as already mentioned,  the way the problem is stated above,   it is actually a QFT  problem in the absence of gravity.    
It is only relevant to gravity when one treats  $\langle 0 | T_{\mu\nu} | 0 \rangle$ as a source in Einstein's equations of General Relativity.    
Thus it would appear that a first step in addressing the vacuum energy density  should focus  on making  mathematical and physical sense of $\langle 0 | T_{\mu\nu} | 0 \rangle$ purely in the context of quantum field theory in flat space,  i.e. without gravity.    This may or may not resolve the cosmological constant problem,   but it is worthwhile exploring  with the theoretical tools we have available at the present time.    
In \cite{AL} we studied this problem for integrable quantum field theory in $d=2$ spacetime dimensions.   Although $d=2$ is considerably simpler,   conceptually the problem is essentially the same as in $4d$ since in $2d$ the calculation \eqref{CC2} also leads to a divergent $\rhovac  \approx \Lambda^2 /4 \pi$.    
We proposed that interactions can actually fix the above simplistic free field calculation.     Using integrability,   we were able to exactly calculate $\rhovac$ 
for a wide variety of models,  including  massive and massless,    and some with and without SSB.     The main point is that it is physically meaningful and calculable without quantum gravity.       It was found that 
for all these models 
\beq
\label{rhovac11}
\rhovac = - \frac{m^2} {2 \, \gcal} 
\eeq
{\it exactly},   where $m$ is a physically measurable  mass scale and $\gcal$ an interaction coupling.   
The main tool that led to this result was Zamolodchikov's analysis of the Thermodynamic Bethe Ansatz (TBA) \cite{ZamoTBA,KlassenMelzer,MussardoBook},   which is a relativistic generalization of 
Yang-Yang thermodynamics \cite{YangYang}.   For many additional references which deal with  some specific models,   we refer to \cite{AL}.   
For the massive case,   in the formula \eqref{rhovac11} $m = m_1$ which is the {\it physical} mass of the {\it lightest} particle and $\gcal$ is a generalized coupling 
which is a  trigonometric sum over certain resonance angles of the 
exact 2-body S-matrix for the scattering of this lightest particle with itself.    See for example \eqref{rhovacexact}  below.  
 This is ultimately a consequence of the S-matrix bootstrap,   which in principle applies 
in all spacetime dimensions. 
For massless cases,   which are renormalization group flows between two conformal field theories,    $m$ can be the  scale of SSB.  
We should add that although we have not considered QFT in general curved spacetime,   the TBA formalism that led to \eqref{rhovac11} does involve QFT on a cylinder,   not flat space.

The above $2d$ results led  us to suggest \cite{AL} that in $4d$,   
\beq
\label{rhovac4d}
\rhovac = -  \frac{m^4} {2 \, \gcal} ~. 
\eeq
In \cite{AL} 
we did not attempt to justify the above $4d$ proposal  in any particular model.     In this paper we will do so for $\lambda \phi^4$ theory.     
We were encouraged to undertake this study  by some recent results 
 from a very different approach involving charged black holes and the notion of a Swampland 
 \cite{Montero1,Montero2}.    There it  was proposed that 
\beq
\label{Montero}
\rhovac <    \frac{m^4}{2 e^2}
\eeq
where $m$ is the mass of a charged particle,  and $\alpha = e^2/4 \pi  $  is the electromagnetic  fine structure constant.      This is weaker than \eqref{rhovac4d} since it is an upper bound rather than an equality.   Remarkably  this  is consistent with \eqref{rhovac4d} if $m$ in \eqref{Montero} is the lightest mass particle and $<$ is replaced with $\leq$. 
In other words the novelty of our proposal \eqref{rhovac4d} is that whereas it is consistent with \eqref{Montero} if $m$ is the lightest mass,  it proposes that the lightest mass particle saturates the inequality leading to an equality.  
One intriguing aspect of \eqref{rhovac4d} is that if $m$ is for  the lightest mass particle and $\gcal \approx 1$,   then the astrophysically measured value of $\rhovac \approx 
10^{-9} {\rm Joule}/{\rm meter}^3$ implies the lightest particle has a mass  on the order of the expected neutrino masses ($0.03 {\rm eV}$).\footnote{Astronomical data is based on WMAP \cite{WMAP}.    The subject of neutrino masses is reviewed in \cite{neutrino}. }

The main goal of this paper is to understand how to obtain  \eqref{rhovac4d}  {\it without} relying on integrability,   at least in some approximation.   
We will also demonstrate that a QFT can have a well-defined cosmological constant even in the absence of spontaneous symmetry breaking.       
     First of all there is no integrability in $4d$ and thus no TBA.     Secondly,   
in the TBA the theory lives on an infinite cylinder of circumference $\beta$;   in thermal field theory $\beta = 1/T$ where $T$ is the temperature.   
In \cite{AL}  we proposed that the cosmological constant  $\rhovac$ is the $\beta$ independent term in the free energy density,   however in the TBA this term is sometimes tricky to extract 
since it can mix with terms coming from conformal perturbation theory.      On the other hand,   it should be possible to compute 
$\rhovac$ directly in the zero temperature quantum field theory,    and this paper shows how to do this for a simple model,   namely the $\lambda \phi^4$ theory,  in a weak coupling approximation.    
We chose to study the latter theory since  this 
alternative calculation   can be compared with exact results for the sinh-Gordon model at small coupling as a check of the method.

In the next section we review the exact $\rhovac$ for the sinh-Gordon model which was originally obtained with the help of the TBA.    We show how this result can be obtained
at weak coupling from a relatively simple calculation without introducing $\beta$ and the TBA\footnote{This short article may thus be viewed as an addendum to \cite{AL}}.    
We then apply this approach to $\lambda \, \phi^4$ theory in $d$ spacetime dimensions and show how to obtain both \eqref{rhovac11},\eqref{rhovac4d}.      
   An interesting feature is that in order to obtain the correct result one must 
analytically continue in $m^2$ from a regime where $m^2$ is negative and has SSB to a physical region with no SSB,   since there is no SSB in the sinh-Gordon model.\footnote{\blue{Note added after publication: 
This need for analytic continuation can be explained by the fact that in the thermodynamic ansatz in 2 spacetime dimensions,   the scalar sinh-Gordon field is  necessarily treated as a fermion.} }   
We will derive a Callan-Symanzik for $\rhovac$ based on the renormalization group for the coupling $\lambda$,   which leads to an RG flow for $\gcal$.  
The two main cases correspond to whether $\gcal$ is marginally relevant or irrelevant.     For the marginally relevant case the cosmological constant {\it decreases} in the flow to low energies.

\section{Generalities for a scalar field in any spacetime dimension}

In this article,  we only consider models of a single scalar field in $d$ spacetime dimensions.   
The classical theory can be defined by the action in euclidean space 
\beq
\label{Vaction}
\CS = \int d^d x \( \smallhalf (\d_\mu \phi)^2 + V(\phi)  \) .
\eeq
As usual we consider the partition function 
$Z = {\rm Tr} \, e^{-\beta H} $ where $\beta$ is the inverse temperature.    
From $Z$ we can calculate the  free energy density $\CF$,   energy density $\CE$,  and pressure $p$ in the usual manner  
\beq
\label{Fp}
\CF =-p =  - \inv{\beta \vol} \log Z ,  ~~~~~~~~~
\CE = - \inv{\vol}  \frac{\d \log Z}{\d \beta} 
\eeq
where $\CV$ is the $d-1$ dimensional spatial volume.   
For arbitrary $\beta$ the above equations determine an equation of state relating $\CE$ and $p$,   which generally does not correspond to a cosmological constant.   
However in \cite{AL} it was shown that the $\beta$ independent term in $\CF$ does correspond to a cosmological constant.    Let us show this here in a different manner.  
First of all consider an arbitrary shift of $V(\phi)$ by a constant $v$,  $V(\phi) \to V(\phi) + v$.      Whereas $Z$ depends on $v$,   correlation functions do not,  since $v$ cancels in 
$\langle \CO \rangle = (\, \int D \phi \, e^{-S} \, \CO \,)/Z$.   

Let us calculate $\rhovac$ in a saddle point approximation.    In the vacuum $\phi$ has no dependence on spacetime   so we can ignore the $\d \phi$ terms.   
The saddle point is then the value of $\phi = \phinot$ satisfying 
\beq
\label{saddle0}
\frac{d V(\phi)}{d\phi}  \Big |_{\phi = \phinot} = 0.
\eeq
The action is then 
\beq
\label{saddle}
\CS_0  =  \int d^d x \, V(\phinot ) = \vol \, \beta \, V(\phinot )  ~~~\Longrightarrow ~~~ Z \approx e^{- \vol \beta \, V(\phinot)}, 
\eeq
since in thermal field theory, euclidean time is a circle of circumference $\beta$.  
This implies a $\beta$ independent free energy density 
\beq
 \CF = V(\phinot ).
\eeq
The equation of state  corresponds to a cosmological constant \eqref{CC1} since it implies the equation of state $\CE = - p$: 
\beq
\label{saddle3} 
\CE = V(\phinot), ~~~~~ p = - V(\phinot).
\eeq
We adopt the standard convention that a positive $\CE$ corresponds to negative pressure $p$:
\beq
\label{saddle4}
\rhovac =   V(\phinot) 
\eeq
in this approximation.

\section{The 2d sinh-Gordon model at weak coupling} 

The sinh-Gordon model is perhaps the simplest integrable  and relativistic quantum field theory.    It can be defined by the action 
\beq
\label{shGaction}
\CS = \int d^2 x \( \inv{8 \pi} (\d_\mu \phi \, \d^\mu \phi ) + 2 \lambdahat  \,  \cosh ( \sqrt{2} \, b \phi ) \) .
\eeq
The $1/8 \pi$ normalization of the kinetic term is such that the two point function has the standard $2d$ conformal field theory normalization:
$\langle \phi (x) \phi(0) \rangle = - \log x^2 $ when $\lambdahat =0$.     The operator $\cosh (\sqrt{2} b \phi )$ is then strongly relevant with scaling dimension 
$-2 b^2$.   
The spectrum consists of a single particle of mass $m$.   Parameterizing the energy and momentum of a particle in terms of a rapidity $\theta$,   
\beq
\label{rapidity}
E = m \cosh \theta , ~~~~~~ p = m \sinh \theta,
\eeq
the exact 2-body S-matrix is 
\beq
\label{shGSmatrix}
 S(\theta)  = \frac{ \sinh \theta - i \sin \pi \gamma }{\sinh \theta  + i \sin \pi \gamma},  ~~~~~ \gamma \equiv  \frac{b^2}{1 + b^2} .
 \eeq

As explained in \cite{AL},   the strict $2d$ analog of the  $4d$ cosmological constant corresponds to the so-called bulk term in the effective central charge $c(\beta m)$. 
The latter can extracted from the TBA,  but without some level of difficulty \cite{ZamoTBA,KlassenMelzer,MussardoBook}.       In the TBA one calculates the free energy on a cylinder of circumference $R = \beta = 1/T$,   where $T$ is temperature.   
However the exact result is quite simple: 
\beq
\label{rhovacexact}
\rhovac = \frac{m^2} {8 \, \sin \pi \gamma } .
\eeq
Since this result depends only on S-matrix parameters,   it must be possible to obtain it directly in the zero temperature quantum field theory,  and this is the primary goal of this paper,  since doing so can provide insights into the $4d$ cosmological constant problem.    

Whereas a shift of the potential by a constant $v$ in the last section would appear to shift the saddle point approximation to $\rhovac$,   there is clearly  no room for such a shift of the  above quoted \eqref{rhovacexact} vacuum energy for the sinh-Gordon model on a finite cylinder.  Once given the S-matrix,  the TBA equations are determined,  and the coefficient of the bulk term in the free energy on a cylinder is completely fixed.    It would be nice to understand this better,   however we suspect it is due to the finite circumference  $R$ of the cylinder that is rendering the  problem well defined.     This leads us to propose  the following principle which eliminates the freedom to shift by $v$:    The only contributions to  the stress-energy tensor in Einstein's General Relativity are properties that can be measured in a  flat space laboratory.\footnote{In plain language,  If you can't relate,  you don't gravitate!}    This rather  conservative principle solves  the usual fine-tuning problems.         Also,  it  is consistent with the Casimir effect,   in that only changes in the vacuum energy density as a result of changing a geometric modulus,   for Casimir it is  the separation of the plates,  is measurable,  since it leads to measurable force.      Indeed  the sinh-Gordon result \eqref{rhovacexact}  is measurable in the finite geometry of a circle of circumference $R$.      In fact it can even be derived on a lattice
\cite{DestriDeVega}.    On the other hand the shift by $v$ is in fact not measurable in flat space by any means whatsoever without gravity. 
We also wish to repeat that the TBA calculations that lead to \eqref{rhovacexact} require studying the theory on a cylinder,   which is not flat spacetime.   

  At small coupling $b$ one has 
\beq
\label{smallb}
\lim_{b \to 0} \, \rhovac = \frac{m^2}{8 \pi b^2} .
\eeq
Note that as the couplling $b \to 0$,   this is a free field limit,  and there is indeed a divergence,   which is consistent with \eqref{CC1}. 
This can be obtained in a simple way using results of the last section.   
The saddle point satisfying \eqref{saddle0} is simply $\phi_0 = 0$,   thus 
\beq
\label{shG2}
\rhovac = 2 \lambdahat .   
\eeq
The above result does not rely on integrability,   and is not exact except  in the $b \to 0$ limit.  
If one allows results from integrability,   then the relation between $\lambdahat$ and the physical mass $m$ and coupling constant $b$ is known exactly \cite{ZamoShG}.  
Since the $\cosh$ potential has dimension $- 2 b^2$,  the scaling dimension of $\mu$ is $2 + 2 b^2$,    thus 
$\mu \propto m^{2 + 2 b^2}$ where $m$ is the renormalized physical mass.     The exact relation is 
\beq
\label{shGmu}
\lambdahat   = \inv{\pi}  \frac{\Gamma (1 - b^2 )}{\Gamma ( b^2 )} \, \[ m \CZ (\gamma) \]^{2 + 2 b^2} , ~~~~~{\rm with} ~~~
\CZ (\gamma)  = \inv{8 \sqrt{\pi} }  \,  \gamma^\gamma \, (1-\gamma)^{1- \gamma} \, \Gamma\( \tfrac{1-\gamma}{2} \) \,  \Gamma \( \tfrac{\gamma}{2} \).
\eeq
In the limit $b^2 \to 0$,  
$\CZ \approx 1/4 b^2$ which implies 
\beq
\label{muapprox}
\lambdahat \approx \frac{m^2}{16 \pi \, b^2},
\eeq
   and this combined with \eqref{shG2} gives the correct limit \eqref{smallb}.  

In the $b\to 0$ limit,  the result  \eqref{muapprox} can be obtained in a much simpler  way without using integrability and this will be useful in the sequel.  
Expanding the $\cosh$ and redefining $\phi \to \sqrt{4 \pi} \phi$,  the lagrangian is 
\beq
\label{cosh4} 
\CL =  \inv{2} (\d_\mu \phi)^2  + \frac{m^2}{2} \phi^2 +  \frac{\lambda}{4!} \phi^4  +  O(\phi^6 ) ,   ~~~ {\rm with } ~~~m^2 = 16 \pi b^2 \, \mu, ~~~~~\lambda = 128 \pi^2 b^4 \, \mu .
\eeq
This naturally leads us to the next section where we consider the cosmological constant for $\lambda  \phi^4$ theory in $d$ spacetime dimensions in light of the above understanding.

\section{ $\lambda  \phi^4 $ theory in $d$ spacetime dimensions.}

The theory is defined by the euclidean action 
\beq
\label{phi4action}
\CS = \int d^d x \,  \(  \inv{2} (\d_\mu \phi)^2  + \frac{m^2}{2}  \phi^2 +  \frac{\lambda}{4!} \, \phi^4 \).
\eeq
Let $\dim{X}$ denote the scaling dimension of $X$ in mass units.  
The classical,  engineering,  dimensions are 
\beq
\label{dims} 
\dim{m} =1 , ~~~~~ \dim{ \phi } = (d-2)/2,  ~~~~~ \dim{\lambda} =  4-d, ~~~~~\dim{\rhovac} = d .
\eeq

\subsection{Saddle point approximation} 

The saddle point equation leads to 
\beq
\label{phi0}
\phinot^2 = - 6 \,  \frac{m^2}{\lambda} ~~~~~\Longrightarrow ~~~ \rhovac = V(\phinot) = - \frac{3}{2} \, \frac{m^4}{\lambda} .
\eeq
As is well known,   a  non-zero real solution $\phinot$ only exists if $m^2$ is negative,    and  there is spontaneous symmetry breaking of the 
$\phi \to - \phi$ symmetry.   Although well-known and simple,  what will be new is how to explain the exact result of the sinh-Gordon model from it.  It is important to note that in the small $b$ approximation to the sinh-Gordon model \eqref{cosh4},  $m^2$ is positive and there is no spontaneous symmetry breaking,   but nevertheless it has a {\it positive} cosmological constant.    As we will argue below,     in order to explain  the $2d$ result \eqref{smallb} we will  need to analytically continue $m^2$ from negative to positive values.\footnote{Equation \eqref{phi0} together with the $\lambda \phi^4$ approximation to the sinh-Gordon model \eqref{cosh4}
leads to $\rhovac = - 3 \mu$ rather than $\rhovac = 2 \mu$ in \eqref{shG2},  however this is clearly due to the approximation of the $\cosh$ potential with a $\lambda \phi^4$ theory.}

Based on the engineering dimensions \eqref{dims} let us define a dimensionless coupling $\gcal$ as follows:
\beq
\label{lambdagcal}
\lambda \equiv  3\,  m^{4-d} \, \gcal ,
\eeq
where by definition $m$ is the true physical mass.    
The above equation is  analogous to  the exact  sinh-Gordon result \eqref{shGmu}.     
Then $\rhovac$ has the desired form stated in the Introduction for any spacetime dimension $d$:
\beq
\label{rhovacd} 
\rhovac = -  \frac{m^d}{2 \gcal} .
\eeq
One sees that for the saddle point approximation to $\rhovac$ in Section II,  
 the main features of the exact sinh-Gordon result at small $b$, including overall factors,  is obtained  if one analytically continues  $m^2 \to -m^2$ which makes $\rhovac$ positive,   and identifies $\gcal = 4 \pi b^2$.      The need to analytically continue in $m^2$ in order to obtain a positive cosmological constant  is not completely clear,  however what  is clear  is that this  is what one needs to do to obtain the correct sinh-Gordon result from \eqref{cosh4}  since the $m^2$ has the wrong sign for there to be a non-trivial $\phi_0$.

%For the small $b$ approximation to the sinh-Gordon model,   based on \eqref{phi0} one finds $\rhovac = -3 \mu = -3 m^2/16 \pi b^2$  rather than $2 \mu$ in \eqref{shG2},  however this  %3 verses 2 is due to 
%he $\lambda \phi^4$ approximation to the sinh-Gordon $\cosh$ potential.     

\subsection{Renormalization group considerations}

The saddle point approximation to $\rhovac$,   namely \eqref{rhovacd},   is not a renormalization group  (RG) invariant.   For the $2d$ sinh-Gordon model,   with a proper RG prescription,  $b^2$ can be viewed as an  RG invariant.     In other dimensions,    $\gcal$ has a non-trivial RG flow,   and one needs to investigate the implications of this.   
Renormalization of $\lambda \phi^4$ theory is well understood (see for instance \cite{Peskin})  however its implications for $\rhovac$ have not been considered previously in much detail, at least to our knowledge.    Being related to a correlation function \eqref{CC1},   $\rhovac$ satisfies a RG differential equation.   This involves absorbing divergences into the parameters $m, \lambda$ and the normalization of the field $\phi$,  which necessarily introduces an arbitrary mass scale $M$,   and a specific renormalization prescription which defines physical parameters,   such as the actual physical mass of particles.        Being a 1-point correlation function which is independent of spacetime coordinates,   these RG equations for $\rhovac$ are simpler than for 
general correlation  functions.    For our purposes,   we want $m$ in \eqref{rhovacd} to be the {\it physical},  measurable mass of a particle.  
For this reason,   the Callan-Symanzik form of the RG equation is most suitable,   since there  the arbitrary renormalization scale $M$ is the actual physical mass $m$.   
In this prescription,   $m$ has dimension $1$ with no anomalous corrections\footnote{$\gamma_m =0$ in the notation in \cite{Peskin}},   and the beta function $\beta_\lambda$  for the coupling $\lambda$ only depends on $\lambda$ and not
$m$.     This  RG equation is 
\beq
\label{Callan} 
\( m \frac{\d}{\d m} + \beta_\lambda   \frac{\d}{\d \lambda } \) \rhovac = \Gamma_\rho \, \rhovac
\eeq
where $\beta_\lambda = m \d_m \lambda$, 
 $\Gammarho$ is the scaling dimension of $\rhovac$,  and 
\beq
\label{betalamba}
\beta_\lambda  (\lambda) = (4-d) \lambda  + O(\lambda^2 ). 
\eeq
Indeed   $\rhovac \propto m^4/\lambda$ as in \eqref{phi0} satisfies the above equation to lowest order with $\Gammarho = d + O(\lambda)$.     
However the higher order corrections to $\beta_\lambda$ imply that the beta function for the classically dimensionless $\gcal$ is non-zero, and the Callan-Symanzik equation now is 
\beq
\label{Callang} 
\( m \frac{\d}{\d m} + \beta (\gcal)   \frac{\d}{\d \gcal } \) \rhovac = \Gammarho  \, \rhovac, ~~~~~ \beta( \gcal) \equiv m \frac{\d \gcal }{\d m} .
\eeq
This  is consistent with $\rhovac \propto m^d /\gcal$ and $\beta (\gcal) =0$ classically.      

Quantum corrections to 1-loop  are known \cite{Peskin} 
\beq
\label{betag}
\beta(\gcal ) = m \frac{d\gcal }{dm} = - \frac{9}{16 \pi^2} \, \gcal^2 + O(\gcal^3) .
\eeq
%which implies $\Gammarho = d - \beta(\gcal)/g $.  

The RG flow toward low energy corresponds to increasing $m$.     Let us fix $\gcal = \gcal_0$ at some high energy scale $m_0$ such as the Planck scale.    Then integrating the one-loop $\beta$ function 
\eqref{betag} one has 
\beq
\label{gofm}
\gcal (m) = \frac{\gcal_0 }{1 + \tfrac{9}{16 \pi^2} \, \gcal_0  \, \log (m/m_0 ) } . 
\eeq
In any spacetime dimension $d$ there are essentially two generic cases to consider:

\bigskip\bigskip

\n  {\bf Marginally irrelevant.} ~~ Here $\gcal_0 >0 $,   and $\rhovac$ is negative.    In the flow to low energies (increasing $m$),  $\gcal \to 0$  and   $\rhovac \to - \infty$. 

\bigskip\bigskip

\n  {\bf Marginally relevant.} ~~ Here $\gcal_0 < 0 $,   and $\rhovac$ is positive.     In the flow to low energies,    $|\gcal |$ increases and  $\rhovac$  slowly flows to $\rhovac=0$ and reaches 
there at 
\beq
\label{mmo} 
m/m_0 = e^{-16 \pi^2/9 \gcal_0} > 1,
\eeq then it changes sign.    
The expononential in \eqref{mmo} implies there can be a very large hierarchy of scales relating the cosmological constant in the UV and IR.

  \bigskip\bigskip
\n There are some features that specifically depend on the spacetime dimension $d$:

\bigskip

\n {\bf $d=2$.}   ~~   Here $\rhovac = - m^2/2 \gcal$.     Recall that for the sinh-Gordon model,   $\rhovac$ is positive and there is no spontaneous symmetry breaking.       Thus in order to reproduce the known exact result in
the sinh-Gordon model at weak coupling,   one must analytically continue $m^2 \to - m^2$ which makes $\rhovac>0$ and is consistent with no spontaneous symmetry breaking, 
i.e. $\phinot =0$.

\bigskip

\n {\bf $d=4$.}    ~~   Here $\rhovac = - m^4 / 2 \gcal$.    Thus the analytic continuation $m^2 \to - m^2$ does not change the sign of $\rhovac$.   A positive cosmological constant requires
a marginally relevant coupling $\gcal$ that is negative.    As explained above,   this can occur for asymptotical free theories in the UV,    where $\gcal \to 0$ and 
$\rhovac \to \infty$ at high energy.

\section{Concluding remarks}

We have argued that the  quantum vacuum expectation value of the stress energy tensor can be well-defined in $d$ spacetime dimensions by including interactions.    The main support for our  analysis is that it can reproduce the exact, small coupling limit for some integrable quantum field theories in $d=2$,  in particular the 
sinh-Gordon model.   This study could provide insight into the cosmological constant problem since the most well-known version of the problem is an issue of QFT in flat space,   
where the source of gravitation is the vacuum expectation value $\langle T_{\mu \nu} \rangle $.      There are other versions of the problem mentioned in the Introduction,  and 
it is not at all clear this is the origin of the observed cosmological constant,  since we have not incorporated gravity.   However  
the problem studied here  is well motivated and posed,   and 
 essentially decouples the problem from classical  and quantum gravity.    

 Based on insights gained in $2d$ we studied the problem for 
$\lambda \phi^4$ theory in $d$ spacetime dimensions and motivated the result $\rhovac = - m^d/2\gcal$ in a saddle point approximation.  This result does not require spontaneous symmetry breaking.           This entails a 
renormalization group equation satisfied by $\rhovac$ which is naturally of Callan-Symanzik type.  For a  marginally relevant coupling $\gcal$,   such as for asymptotically free theories,
$\rhovac$ can flow from large positive values to zero,   and this flow introduces a large hierarchy of energy scales.       

If our analysis proves to be correct,   then there are many open avenues for  exploration.     It would be interesting to try and extend our results to theories with both bosons and fermions as in the Standard Model of particle physics.     In fact, based on our analysis of simpler models,   conceptually the cosmological constant in the Standard Model is {\it in principle} computable,  but difficult;    it is non-perturbative,    and  perhaps can be computed  on a lattice from finite size or temperature effects.   The computation of vacuum energy density based on the TBA described in \cite{AL} is actually a finite size effect since the formalism involves quantum fields on a cylinder.      Indeed,   it was shown how to obtain exact results for the vacuum energy density for models like the sinh-Gordon model from the lattice \cite{DestriDeVega}.   In fact it can in principle be measured in a laboratory  through finite size effects,  as for the usual Casimir effect.     

We have not at all explored  the consequences of including $\rhovac$ in the 
temporal and thermal evolution of the universe;  as already stated we decoupled the cosmological constant problem from gravity itself and thus cosmology.      However we suggested one scenario wherein $\gcal$ is a negative marginally relevant coupling,   for instance  for an asymptotically free theory,   and $\rhovac$ flows to zero at low energies,   indicating a kind of ``cosmic freedom" in that the cosmological constant does not dominate at very  late times.

\section*{Acknowledgments}
 
We  wish to thank Miguel Montero and  Gerben Venken for pointing out their  potentially related work  \cite{Montero1,Montero2} and discussions.

\end{document}